\documentclass[12pt,reqno]{article}
\usepackage{amsmath,hyperref,amssymb}
\usepackage{subfigure,graphicx,url}

\renewcommand{\b}{\bar}
\newcommand{\be}{{\bar{1}}}
\newcommand{\bt}{{\bar{2}}}
\allowdisplaybreaks \numberwithin{equation}{section}

\DeclareSymbolFont{AMSa}{U}{msa}{m}{n} \DeclareSymbolFont{AMSb}{U}{msb}{m}{n}
\DeclareMathSymbol{\fieldR}{\mathalpha}{AMSb}{"52} \DeclareMathOperator{\Li}{Li}

\begin{document} 
\begin{flushright} \small
  arXiv:0801.3949 [hep-th] \\ ITP--UU--08/03 \\ SPIN--08/03
\end{flushright}
\bigskip
\begin{center}
 {\large\bfseries On NS5-brane instantons and volume stabilization}\\[5mm]
Hugo Looyestijn and Stefan Vandoren \\[3mm]
 {\small\slshape
 Institute for Theoretical Physics \emph{and} Spinoza Institute \\
 Utrecht University, 3508 TD Utrecht, The Netherlands \\
 {\upshape\ttfamily h.t.looyestijn, s.vandoren@phys.uu.nl}\\[3mm]}
\end{center}
\vspace{5mm} \hrule\bigskip \centerline{\bfseries Abstract} \medskip
We study general aspects of NS5-brane instantons in relation to the
stabilization of the volume modulus in Calabi-Yau compactifications of
type II strings with fluxes, and their orientifold versions.
These instantons correct the K\"ahler potential and
generically yield significant contributions to the scalar potential at intermediate values of
string coupling constant and volume. Under suitable conditions they
yield uplifting terms that allow for meta--stable de Sitter vacua.

\bigskip
\hrule\bigskip
\section{Introduction}
Flux compactifications and moduli stabilization in string theory
has been an active area of research in recent years. For some
reviews on the topic, see
\cite{Grana:2005jc,Douglas:2006es,Blumenhagen:2006ci}. Of
particular interest is the stabilization of the volume modulus of
the compactification manifold. In the effective supergravity
approximation, one assumes that curvatures are small and hence the
volume of the internal space is large compared to the scale set by
the string length $l_s=2\pi {\sqrt {\alpha'}}$. Stabilization by
finding the minima of the scalar potential of the low-energy
effective action must therefore lead to a large value for the
volume for this approach to make sense. This is one of the basic
assumptions in the KKLT scenario \cite{Kachru:2003aw}, or the more
recent so-called ``large volume scenarios'' (LVS)
\cite{Balasubramanian:2005zx,Conlon:2005ki}. The scale set by the
volume has important consequences for low-energy physics, such as
supersymmetry breaking and inflation, see e.g.
\cite{Conlon:2005ki,Kachru:2003sx,Conlon:2005jm,Berg:2007wt,Berg:2004ek}.

For type IIB flux compactifications on Calabi-Yau (CY) manifolds
(and their orientifolds), the volume is stabilized by
nonperturbative effects, such as stringy D3-brane instantons whose
Euclidean worldvolume wraps a four-cycle in the CY. Such an
instanton stabilizes the volume of the corresponding four-cycle
and therefore the associated K\"ahler modulus. The relation
between the four-cycle and two-cycle volumes is known in
principle, but is complicated in practice since it requires
inverting a set of coupled quadratic equations involving the triple
intersection numbers, as we review in the next section. Therefore,
although stabilization of all K\"ahler moduli indeed stabilizes
the entire volume of the CY, it requires a case by case analysis
to fix it at large values\footnote{The situation might seem to
look better in type IIA theories, since there one can stabilize
the K\"ahler moduli directly at the classical level by switching
on fluxes \cite{DeWolfe:2005uu}. However, the K\"ahler moduli are
fixed again by solving a set of quadratic equations involving the
triple intersection numbers, see equation (4.36) in
\cite{DeWolfe:2005uu}. So to find large values for the total
volume, one ends up with similar difficulties as in type IIB.},
see for instance \cite{Berg:2007wt} for a recent analysis.

The fact that a wrapped Euclidean $p$-brane can stabilize the
volume of a $p+1$-cycle naturally raises the question of whether
the overall CY-volume can be stabilized directly by wrapping a
Euclidean fivebrane over the entire CY. From this point of view,
one could expect NS5-brane instantons to play an important role in
relation to volume stabilization~\footnote{In $N=2$
compactifications of type IIB, one also expect D5-brane instantons
to contribute. However, after orientifold projection with O3/O7 planes, the D5 brane is not BPS and therefore harder to analyze. Furthermore, we look for a mechanism that also applies to IIA and heterotic string theories.}. In case such instantons contribute to the
low-energy scalar potential, they will do so with exponentially
suppressed terms of the form
\begin{equation}\label{NS5-inst}
\exp\,[-S_{{\rm NS}5}\,]=\exp\left[-\frac{{\cal V}}{g_s^2}\,\right]\ ,
\end{equation}
where $S_{{\rm NS}5}$ is the (real part of the) one-instanton action,
${\cal V}$ is the volume of the CY in dimensionless units, and $g_s$ is the
ten-dimensional string coupling
constant. This can be compared to the contribution of a D3-brane instanton,
of the form
\begin{equation}
{\rm exp}\,[-S_{{\rm D3}}\,]=\exp\left[-\frac{{\rm vol}(\gamma_4)}
{g_s}\,\right]\,.
\end{equation}
Here, vol($\gamma_4$) is the volume of the four-cycle $\gamma_4$.

At weak string coupling constant, one expects the D3-brane
instantons to dominate over the NS5-brane instantons, such that
one can safely ignore the latter. The two exponents only are of the
same order of magnitude when
\begin{equation}\label{eq:ordermagnitude}
\frac{{\rm vol}(\gamma_4)}{{\cal V}}=\frac{1}{g_s}\,.
\end{equation}
For small four-cycles and weak string coupling, $g_s \leq 1$, this is never
satisfied. However, there are plenty of CY manifolds which have four-cycles
with larger volume than the total volume, as we review in the next section,
so some care is needed to make this argument, especially when $g_s \sim 1$. Similar considerations hold for
compactifications of type IIA strings on CY threefolds, in which membrane
instantons arise by wrapping Euclidean
D2-branes over three-cycles.

There is another reason to be careful in ignoring the fivebrane instantons.
Assuming that both instantons contribute to the scalar potential in the effective action and $\text{vol } \gamma_4 < \mathcal V$, the exponents above can still be multiplied by prefactors to make them of the same order, especially at intermediate
string coupling $g_s \sim 1$. We will show this more explicitly in the next section,
for values ${\cal V}\approx 100$ (which are the typical values of the
original KKLT approach). The existing vacua of the LVS scenarios at
${\cal V}\approx 10^{13}$ are not effected by NS5-brane instantons. However, at smaller volumes $\mathcal V \sim 100$, additional vacua can arise with
interesting properties.

The purpose of this paper is to analyze the effects of NS5-brane
instantons in relation to the stabilization of the volume modulus.
In particular, we show that under certain conditions, the
contributions from NS5-brane instantons yield uplifting terms in
the scalar potential that can lead to meta--stable de Sitter vacua.
Most of the work on moduli stabilization has focused on $N=1$ supersymmetry in four
dimensions, as they give rise to semi-realistic string vacua. In
such models, like e.g. type IIB strings on Calabi-Yau
orientifolds, moduli can be stabilized by combining the effects of
fluxes and quantum corrections coming from perturbative
corrections to the K\"ahler potential and D3-brane instanton
corrections to the superpotential. However, the K\"ahler potential
is subject to higher loop corrections in $\alpha'$ and $g_s$ which
are not known explicitly. For a recent discussion on this, see
\cite{Berg:2007wt, Berg:2005yu,  Cicoli:2007xp}. As we will show, the
nonperturbative corrections of the form \eqref{NS5-inst} also
contribute to the K\"ahler potential, and are generically
subleading with respect to the first perturbative
corrections, but could compete with next--to--leading perturbative
corrections. For this reason, our investigations are more
meaningful in $N=2$ models, since in that case higher order
corrections are absent due to the constraints from
$N=2$ supersymmetry. This fact also
motivated the authors of \cite{Kachru:2004jr} to study $N=2$
moduli potentials in type IIA flux compactifications. These toy
models can serve as good approximations for the more realistic
$N=1$ string vacua. Moreover, for $N=2$ theories in type IIA,
there are some explicit results known about the contribution of
NS5-brane instantons
\cite{Alexandrov:2006hx,Davidse:2004gg,deVroome:2006xu,
Gutperle:2000ve} to the effective action for the hypermultiplets.

The plan of the paper is as follows. In section 2, we present the
generic form of an NS5--brane instanton correction to the scalar
potential. We study this in the setting of $N=1$ supergravity in
four dimensions, and investigate the relation with the KKLT and LVS
scenarios. In section 3, we discuss IIA strings compactified on a
(rigid) CY, for which there is some explicit knowledge on
NS5-brane instantons. We investigate the stability of the volume,
and find the possibility that NS5-brane instantons can produce de
Sitter vacua. We then truncate this model preserving local $N=1$
supersymmetry, and determine the K\"ahler and superpotential.

\section{Volume stabilization}
In this section, we review certain aspects of the KKLT scenario and
discuss some of the subtleties that can arise in
stabilizing the volume at large values in IIB orientifold compacti\-fications.
We then include terms that
mimic the contributions from NS5-brane instantons to the K\"ahler
potential, and re-analyze the stabilization of the volume modulus.

We consider an orientifold of type IIB string theory on a CY 3-fold with O3/O7 planes.
The cohomology groups $H^{(p,q)}$ are split under the orientifold mapping into odd and even forms, and hence their dimensions
split as $h^{p,q} = h_+^{p,q} + h_-^{p,q}$. We follow the notation of~\cite{Grimm:2004uq}, although we change a few names and numerical factors.

Let us first list the various chiral fields. The field $\tau$ contains the axion and dilaton, and
is defined by $\tau = l + i{\rm e}^{-\phi_{10}}$. The fields $T_i$ are defined by
\begin{equation}\label{chiral-fields1}
T_i = \tau_i + i h_i - 2 \zeta_i,\quad i,j=1,\ldots,h^{1,1}_+,
\end{equation}
where
\begin{equation}\label{chiral-fields2}
\zeta_j = - \frac i {2 (\tau - \bar \tau)} d_{j a b}G^a (G-\bar G)^b,\quad  G^a = c^a - \tau
b^a,\quad a,b=1,\ldots,h^{1,1}_-.
\end{equation}
The $\tau_i$ capture the sizes of the even four--cycles under the orientifold projection and $h_i$
are real field that arise by expanding the $C_4$ gauge field over these four--cycles. The fields $b^a, c^a$ are the
expansions of the $B_2$ and $C_2$ forms respectively over the $h^{1,1}_-$ cycles. Notice that the
definition of $\zeta_i$ contains intersection numbers of even and odd two--cycles.

The four--cycles $\tau_i$ are related to the two--cycles $t^i$ by the triple intersection numbers
$d_{ijk}$ as
\begin{equation}\label{2-4cycles} \tau_i = d_{ijk}t^jt^k\,.
\end{equation}

The total volume $\mathcal V$ is only implicitly known as a function of the $N=1$ chiral
coordinates through the relation
\begin{equation*}
   \mathcal V = \frac 1 6 d_{ijk}t^it^jt^k.
\end{equation*}
To write the volume in terms of the chiral fields we first use  the definitions~\eqref{chiral-fields1},~\eqref{chiral-fields2}
to find
\begin{align*}
  \tau_i = \frac 12 (T_i + \bar T_i) -\frac i 2 \, d_{iab}b^ab^b (\tau - \bar \tau)
\end{align*}
or in terms of the chiral fields
\begin{align*}
  \tau_i = \frac 12 (T_i + \bar T_i) -\frac i 2 \frac {1}{(\tau - \bar \tau)}\, d_{iab} (G-\bar G)^a (G-\bar G)^b.
\end{align*}
One then solves the quadratic equations in equation~\eqref{2-4cycles} to obtain functions $t^i(\tau_j)$,
and one obtains the volume $\mathcal V$ depending on the chiral fields via $\{\tau-\bar \tau,T_i + \bar T_i,(G-\bar G)^a\}$.

The type IIB K\"ahler potential is given by~\cite{Becker:2002nn,Grimm:2004uq}
\begin{align}\label{KIIBLVS}
  K &= K_{\rm cs}(U,\bar U) + K_{\rm k}(\tau,T,G)\\
  K_{\rm k}&=-\ln\left[ -i (\tau - \bar \tau) \right] - 2 \ln \left[\mathcal V(\tau,T,G)+ \xi\,
  \text{Im}(\tau)^{3/2}\right].\label{KIIBLVS23}
\end{align}

The K\"ahler potential $K_{\rm k}$ in \eqref{KIIBLVS23} is the tree-level expression, together with the
leading perturbative $\alpha'$ correction proportional to the parameter $\xi =- \frac
{\chi(CY)\zeta(3)}{2(2\pi)^3}$, containing the Euler number $\chi(CY)$ of the internal Calabi-Yau $M$ (we use conventions where $l_s = 2 \pi \sqrt {\alpha'})$. The complex
structure deformations $U$ are described by $K_{\text {cs}}$, whose precise form is not important.
Higher string loop corrections could give a dilaton--dependence to $K_{\rm cs}(U,\bar U)$, but this is beyond the approximation we are working in.

The scalar potential for a K\"ahler potential $K$ and superpotential $W$ is given by
\begin{equation}\label{scalarpotential}
V={\rm e}^K\Big(K^{\alpha\bar \beta}D_\alpha WD_{\bar \beta}{\overline W}-3|W|^2 \Big),
\end{equation}
where the indices $\alpha,\bar \beta $ run over all chiral fields, with $K^{\alpha\bar\beta}$ the
inverse K\"ahler metric.

\subsection{Fluxes and D3--brane instantons}
The tree--level superpotential is given by
\begin{align}\label{d3superpotential}
  W = W_0(\tau,U)\ = \int_M \Omega \wedge G_3\,,
\end{align}
where $G_3$ is the complex combination of the background three--form fluxes, given by $G_3 = F_3 -
\tau H_3$.

We now assume that the complex structure moduli $U$ and the axio-dilaton $\tau$ are stabilized at a
higher energy scale at a SUSY minimum, by demanding $D_\tau W=D_UW=0$. To stabilize the K\"ahler
moduli we add the nonperturbative instanton corrections of wrapping Euclidean D3--branes over four--cycles. The superpotential
is then given by~\cite{Kachru:2003aw}.
\begin{align}\label{WIIB}
  W = W_0 + \sum_i A_i {\rm e}^{-a_i T_i}\,,
\end{align}
where $A_i$ and $a_i$ are treated as field--independent parameters. In the literature one often considers the case where the
$G^a$ fields are absent. Using the expressions~\eqref{KIIBLVS} and~\eqref{WIIB}, one finds the following scalar
potential \cite{Becker:2002nn,Balasubramanian:2004uy}
\begin{multline}\label{VLVST}
V = {\rm e}^K \Bigg[ K^{j{\bar k}} \left( a_j A_j a_k \bar A_k {\rm e}^{-a_jT_j-a_k\overline{ T_k}} -
\left( a_j A_j {\rm e}^{-a_jT_j} \overline W K_{\bar k} +\text{c.c.} \right) + K_j K_{\bar k}
|W|^2 \right) -3 |W|^2 \Bigg].
\end{multline}
where $K^{i\bar \jmath}$ are the components of the inverse metric $K^{\alpha\bar\beta}$ in the
directions of the K\"ahler moduli. In the absence of the $G^a$ fields, there is no--scale structure at tree--level, leading to $K^{i\bar \jmath} K_i K_{\bar \jmath} = 3$. This no--scale structure is broken when $\alpha'$--corrections are included, and one then finds (see~\cite{Becker:2002nn}, or for some further details of the calculation, see appendix~\ref{ap:ns5calc})
\begin{multline}\label{VLVS}
V = {\rm e}^K \Bigg[ K^{j{\bar k}}\left( a_j A_j a_k \bar A_k {\rm e}^{-a_jT_j-a_k\overline {T_k}} -
\left( a_j A_j {\rm e}^{-a_jT_j} \overline W K_{\bar k} +\text{c.c.} \right) \right) + 3 \xi
\frac{\xi^2 + 7 \xi \mathcal V + \mathcal V^2}{(\mathcal V-\xi)(2\mathcal V+\xi)^2} |W|^2 \Bigg].
\end{multline}

The various studies (see for example \cite{AbdusSalam:2007pm}) of
these potential indicate a large volume AdS vacuum,
which can be realized in explicit models. For example, in the
$\mathbb P^4_{[1,1,1,6,9]}$ model, which yields two K\"ahler moduli, the volume is expressed in terms of
the 4--cycle volumes $\tau_s, \tau_b$ as
\begin{align}\label{eq:volume_two_kahler}
  \mathcal V = \frac 1 {9\sqrt 2} \left( \tau_b^{3/2} -
  \tau_s^{3/2} \right).
\end{align}
This already gives an example where a four--cycle volume can be bigger then the total volume, as mentioned below~\eqref{eq:ordermagnitude}\footnote{Take for example the values $\tau_s \sim 4.6$ and $\tau_b \sim 120$ which give a total volume $\mathcal V \sim 100$.}.

To remain in the regime of a geometrical compactification, one needs $\tau_b >
\tau_s$. To obtain a large volume, one needs to arrange $\tau_b \gg \tau_s$. This is then solved
self--consistently: one assumes a cycle to be small and the other large, approximates the potential
in this regime and then searches for sets of vacua. In such a two--K\"ahler model this is doable,
but a more general model will have many different K\"ahler moduli. To express the volume in terms
of the four-cycles, one first has to solve the system of many coupled quadratic equations
\eqref{2-4cycles} and use the explicit form of the triple intersection numbers. Even for simple models with a few K\"ahler moduli, this can lead to equations which are not solvable analytically. Numerical methods can be used, but have their own limitations. One then has to find a limit on the four--cycles that leads to a large volume ${\cal V}$. This will be very difficult without any analytical control. Overall, it seems
desirable to have a different mechanism that stabilizes the volume at once, without the need to
stabilize the individual cycles that build up the total volume. A prime candidate for such a
mechanism is the NS5-brane instanton, to which we turn now.

\subsection{Adding NS5-brane instantons}

We now add a correction due to the wrapping of an Euclidean NS5--brane over the entire CY. Such an
instanton configurations contributes to correlators proportional to $\exp(-\mathcal V/g_s^2)$,
where $g_s$ is the 10--dimensional string coupling constant. The volume can not be expressed as a
holomorphic function of the $N=1$ chiral fields. We therefore expect that the NS5--brane does not
correct the superpotential, but instead it will correct the K\"ahler potential,
\begin{align}\label{KIIBNS5}
  K_{\text {NS5}}&=B\, \mathcal V^n \exp(-\mathcal V/g_s^2)\ = B\, \mathcal V^n \exp\left(\frac 14 \mathcal V (\tau - \bar
  \tau)^2\right).
\end{align}

The factor of $\mathcal V^n$ represents the leading power of the instanton measure and the one-loop
determinant of the fluctuations around the instanton solution. There is a proportionality factor
$B$ that could -- in principle -- depend on the moduli $G^a$ and the dilaton $\tau$. We expect no dependence on the complex structure moduli $U$ since the NS5--brane cannot probe the individual 3--cycles. Furthermore, before the orientifold projection, the NS5--branes correct the moduli space of K\"ahler deformations, and not the complex structure deformations. The prefactor
$\mathcal V$ is absent in an instanton corrected superpotential: because the superpotential is a
holomorphic function of the chiral fields $T_i$, any non--trivial function $A_i(T)$ in \eqref{WIIB}
breaks the shift symmetry on the imaginary parts $h_i$ completely and is therefore forbidden. Since
instantons are expected to break the shift symmetries to a discrete subgroup only, we can use
superpotentials of the form $\exp(-a_i T_i)$. The K\"ahler potential does not need to be
holomorphic, and we can therefore only use the real parts of the chiral fields $T_i$, which combine
into powers of the volume $\mathcal V$.

A further argument in favor of \eqref{KIIBNS5} comes from the parent $N=2$ theory. NS5-branes
correct the hypermultiplet moduli space even in the absence of fluxes \cite{Becker:1995kb}. In IIB
compactifications, the hypermultiplet moduli are counted by the K\"ahler moduli, and the kinetic
terms of these scalars receive corrections from NS5-brane instantons. After the orientifold
projection, one expects these corrections to survive, even after fluxes are turned on. Our claim is
that, to leading order, they enter the K\"ahler potential in they way described in \eqref{KIIBNS5}.
We elaborate further on this the next section for type~IIA
compactifications, where we can determine the one--instanton NS5--brane contribution explicitly in some special
cases \cite{Alexandrov:2006hx}.

We have to ask in which regime this approach makes sense. We want to remain in the one--instanton
regime, and not consider multiple instanton contributions since nothing is known about them.
Because an NS5--brane instanton scales as $\exp(-\mathcal V/g_s^2)$, this requires that
\begin{align}\label{eq:cond_ns5}
  \frac {g_s^2}{\mathcal V} < 1\,.
\end{align}
However, since the NS5-brane is the magnetic dual of the fundamental string, one also has to
consider perturbative effects, which we expect to organize an expansion in powers of
$g_s^2/\mathcal V$. Besides that, there are also higher
order $\alpha'$ corrections which are left out. Only the first perturbative correction is included
in our analysis in \eqref{KIIBLVS}. In $N=2$ theories these higher string-loop corrections are
absent~\cite{RoblesLlana:2006ez, Antoniadis:2003sw}, which makes the discussion of the NS5-brane instanton more reliable. We discuss such models
in the next section. We take a pragmatic approach here and isolate the NS5-brane instanton
correction from all other corrections in the K\"ahler potential $K_0$. Hence we write
\begin{equation}
K=K_0(\tau,T,U,G)+K_{{\rm NS5}}(\tau,T,G)\ ,
\end{equation}
and expand to leading order in $K_{{\rm NS5}}$. To obtain the expression for the inverse K\"ahler metric $K^{\alpha \bar \beta}$ (where $\alpha
= \{\tau,T,G,U\}$ lists all the chiral fields) in the direction of the K\"ahler moduli, we use
\begin{align}\label{ns5inverse}
K^{i \bar \jmath} = K_0^{i \bar \jmath} - K_{{\rm NS5}}^{i \bar \jmath}, \quad K_{{\rm NS5}}^{i \bar \jmath} \equiv K_0^{i \bar \alpha} {K_{{\rm NS5}}}_{\bar \alpha \beta} K_0^{\beta \bar \jmath}\,,
\end{align}
and now $K^{i \bar \jmath} K_{\bar \jmath k} = \delta^i_k + \mathcal O(K_{\rm NS5}^2)$.

In principle, there could be a dependence in $K_{{\rm NS5}}^{i \bar \jmath}$ on the fields $b^a$, which enter through the definition of the chiral fields $G^a$ in~\eqref{chiral-fields1} and~\eqref{chiral-fields2}. As explained at the end of appendix~\ref{ap:ns5calc}, the exact dependence on $b^a$ is subleading in $K_{{\rm NS5}}^{i \bar \jmath}$.

To obtain the scalar potential, we work out equation~\eqref{scalarpotential}, setting $D_\tau W=D_UW=0$. For simplicity, we now set $G^a$ to zero. It would be interesting to consider the effects of non--zero $G^a$, but this is beyond the scope of this article. Using this in~\eqref{scalarpotential} leads to
\begin{equation}\label{pert-pot}
V=V_0+V_0K_{{\rm NS5}}-{\rm e}^{K_0}K_{{\rm NS5}}^{i\bar \jmath}\,|D^{(0)}_iW|^2 +{\rm
e}^{K_0}K_0^{i\bar \jmath}\Big(\partial_i K_{{\rm NS5}}\,WD^{(0)}_{\bar \jmath} {\overline W}+{\rm
c.c.}\Big).
\end{equation}
The potential in absence of $K_{\rm NS5}$ is denoted $V_0$, $D^{(0)}_iW=\partial_iW+(\partial_iK_0)W$ and
$K_0^{i\bar \jmath}$ is the inverse of $K_{0,\alpha \bar \beta}$ in the directions of the K\"ahler
moduli. Remind that all chiral fields are labeled by $\alpha,\bar\beta$, and the K\"ahler moduli
are a subsector thereof labeled by $i,\bar \jmath$. Formula \eqref{pert-pot} in fact holds for any
perturbation of the K\"ahler potential, labeled by $K_{{\rm NS5}}$.

{}From equations~\eqref{ns5inverse}--\eqref{pert-pot} we can see that we have to compute ${K_{{\rm
NS5}}}_{\bar \alpha \beta}$, so we have to take derivatives with respect to all chiral fields. This
is to be contrasted to the situation in which one modifies the superpotential: the derivatives with
respect to $\tau$ and $U$ are contained in $D_\tau W$ and $D_UW$, which are set to zero.

Suppose for simplicity that $W$ does not depend on the K\"ahler moduli (such as in
equation~\eqref{d3superpotential}), then $D^{(0)}_iW_0 = K^{(0)}_i W_0$ and hence
\begin{equation}\label{pert-pot-Wcns}
V=V_0+V_0K_{{\rm NS5}}-{\rm e}^{K_0}K_{{\rm NS5}}^{i\bar \jmath}\,|\partial_i K_0|^2 |W_0|^2 +{\rm
e}^{K_0}K_0^{i\bar \jmath}\Big(\partial_i K_{{\rm NS5}}\,\partial_{\bar \jmath} K_0 +  {\rm
c.c.}\Big)|W_0|^2.
\end{equation}

Using expression~\eqref{KIIBLVS} for $K_0$, we find that $K_0^{i\bar \jmath}$ is proportional to
$t^it^j$ and $d^{ij}$ (the inverse of $d_{ij} = d_{ijk}t^k$), and both $\partial_iK_0$ and
$\partial_i K_{\rm NS5}$ are proportional to $d_i = d_{ij}t^j$. Upon contracting indices, this will
combine nicely into powers of $\mathcal V$. The precise calculation can be found in
appendix~\ref{ap:ns5calc}. Ignoring the D3--brane instantons, and other subleading corrections to the K\"ahler potential, the leading correction to $V_0$ is found to be
\begin{align*}
  V_{\text {NS5}} = - \frac 98 B |W_0|^2 g_s^{-3} \mathcal V^n  \exp(-\mathcal V / g_s^2) \equiv \hat B \mathcal V^n  \exp(-\mathcal V / g_s^2).
\end{align*}
In the last expression, we have absorbed all the constants into a new
prefactor $\hat B$.

\subsection{Analysis of the model}

We have motivated the scalar potential
\begin{align}\label{VLVSNS5}
  V = V_0(\mathcal V) + \hat B \mathcal V^n {\rm e}^{-\mathcal V/g_s^2}\,,
\end{align}
where $\hat B, n$ are constants.

A general situation achieved in moduli stabilization is indicated on the left plot below, which displays
an AdS minimum.
\begin{figure}[!h]%
\begin{center}%
\parbox{2.2in}{  \includegraphics[width=5cm]{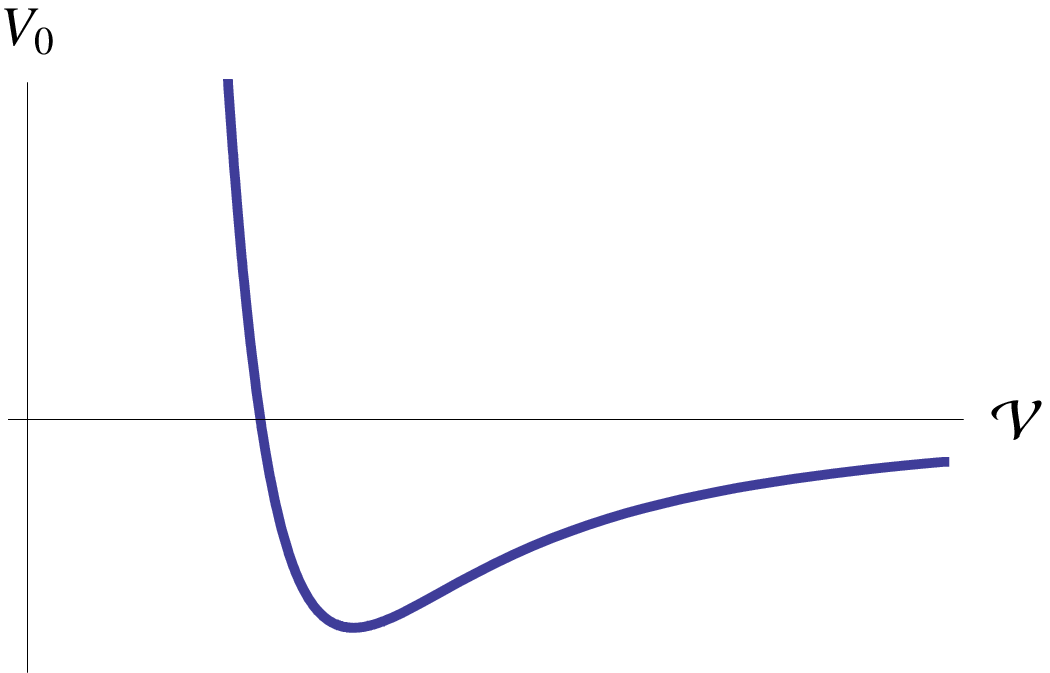}}%
\parbox{2.2in}{  \includegraphics[width=5cm]{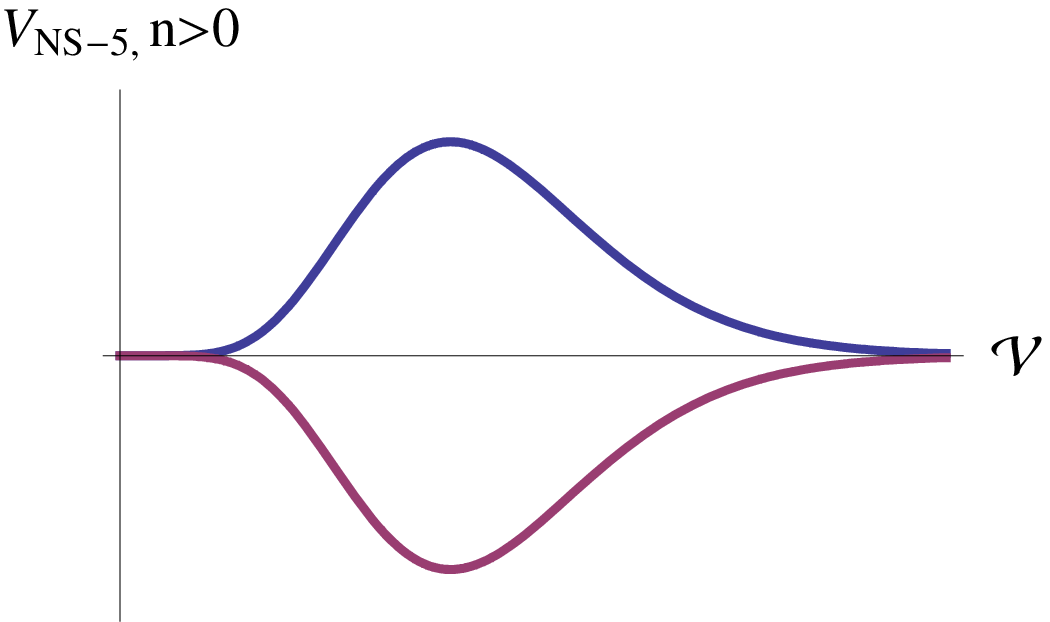}}%
\parbox{2.2in}{  \includegraphics[width=5cm]{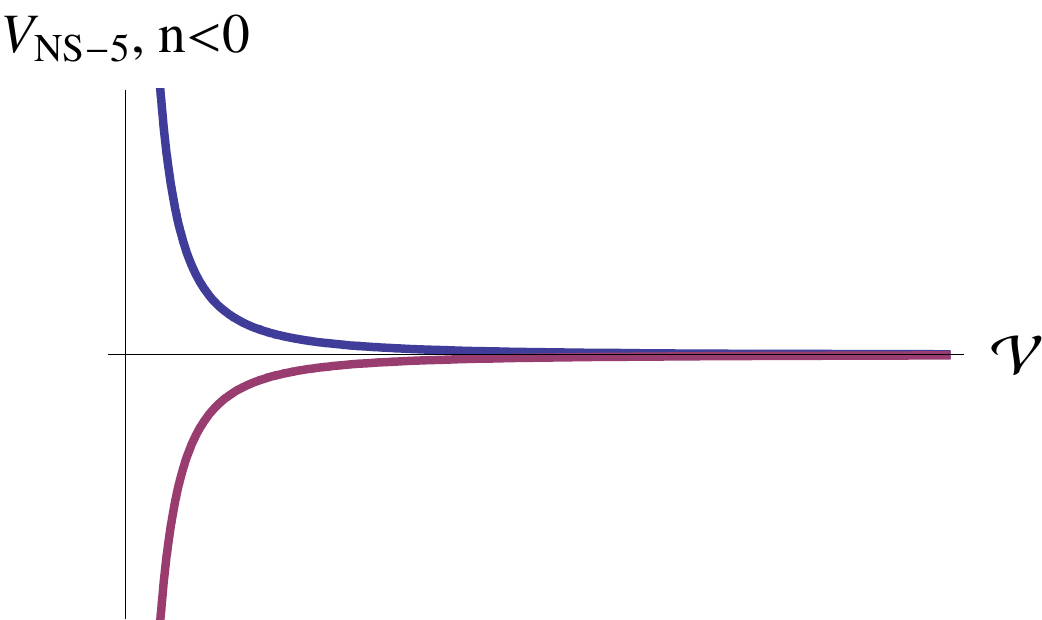}}%
\end{center}%
\caption{General shape of the three terms in the potential. The first is the AdS minimum of
$V_{\text {pert}}$. The second and third are the NS5--brane contribution $\hat B \mathcal V^n
 {\rm e}^{-\mathcal V/g_s^2}$, with $n$ positive and negative, respectively. The possible signs of $\hat
B$ are given by the upper line ($\hat B$ positive) and the lower line ($\hat B$ negative).}
\end{figure}
We will investigate the physics of the NS5--brane contribution, depending on the signs of the
parameters $\hat B$ and $n$. The results are summarized in the table below. The most interesting
case is when $n$ is positive and $\hat B$ is positive (the upper line in the middle graph). The
exact location of the bump depends on the value of $n$. Its value is at this point undetermined, but we will argue in the next section that
\begin{align}\label{eq:valueofn}
n = - 3 - \frac{\chi}{12\pi}\,.
\end{align}
This is the value obtained in the $N=2$ theory, and we assume that its order of magnitude remains the same in the orientifolded $N=1$ theory.

For $n>0$, the contribution from the five--brane is positive and will therefore certainly increase the value of the potential at the minimum. Depending on the strength $\hat B$ and the location of the contribution it can produce new vacua or uplift the
existing minimum to a de Sitter vacuum. With $n$ positive and $\hat B$ negative, the five--brane
yields a negative energy contribution to the potential function. Depending on the parameters, it
will introduce new vacua or lower the potential energy at the location of the existing vacuum. The
situation is also interesting if the existing scenarios do not stabilize the volume and we obtain a
run--away potential which behaves as $V_0 \simeq \mathcal V^m$, with $m$ negative. A five--brane
contribution with positive $n$ can then provide the volume stabilization. The possible signs of $n$ and $\hat B$ and their results are summarized in
the table below.
\begin{center}
\begin{tabular}{|c|c|l|}
  \hline
  $n$ & $\hat B$ & Result \\
  \hline
  + & + & Increased vacuum energy (uplift), with de Sitter vacua \\
  + & -- & Decreased vacuum energy, increased number of minima  \\
  -- & + & Increased vacuum energy (uplift) \\
  -- & -- & Decreased vacuum energy\\ \hline
\end{tabular}
\end{center}

We have outlined the model and the qualitative behavior, and will now investigate the numerics. The
general scenario is described by
\begin{align}\label{ns5correc}
  V = V_0(\mathcal V) + V_{\text {NS5}},\quad  V_{\text {NS5}} = \hat B \mathcal V^n {\rm e}^{-\mathcal V/g_s^2}.
\end{align}
We have not really specified $V_0$ here; one can choose a favorite scenario in which the K\"ahler moduli are stabilized, and then investigate
the influence of the NS5--brane contribution. The expression for $V_{\text {NS5}}$ will, in
general, be more complicated: if we take a non--trivial $W$ (e.g. including D3--brane instantons as
in~\eqref{WIIB}), there will be mixing terms consisting of K\"ahler perturbations and
superpotential perturbation in terms like $|D_i W|^2$. These terms are typically suppressed, or at most of the same order as in~\eqref{ns5correc}. In both cases, the numerical analysis below remains the same.

The scalar potential~\eqref{ns5correc} contains an overall term $\exp(K_{\text {cs}}(U,\bar U))$, whose exact value depends
on the details of the complex structure moduli stabilization. We set this factor to unity to compare with other models~\footnote{In $V_{\rm NS5}$ it was absorbed in the factor $\hat B$.}. The KKLT
scenario stabilizes the volume at $\mathcal V = \mathcal O(100)$ at a minimum of $V_0 = -2 \cdot
10^{-15}$. The LVS on $\mathbb P^4_{[1,1,1,6,9]}$ stabilizes the volume at $\mathcal V = \mathcal
O(10^{12})$ with a value of  $V_0= -6 \cdot 10^{-37}$~\cite{Berg:2007wt}.

Using $B=-10, W_0=1$ (implying $\hat B = 90/8 g_s^{-3}$) and an Euler number of $\chi = -300$, we find (using~\eqref{eq:valueofn}) the values
\begin{equation}
\begin{array}{l|lllll}
  V_{\text {NS5}} & \mathcal V=10 & \mathcal V=13     & \mathcal V=20  & \mathcal V=60  & \mathcal V=130 \\
  \hline
  g_s=0.5 & 10^{-10}& 10^{-15} & 10^{-26} & 10^{-93} & 10^{-213}\\
  g_s=1   & 10^{2} & 10^{0} & 10^{-1} & 10^{-16} & 10^{-44} \\
\end{array}
\end{equation}

For very weakly coupled strings, $g_s < 0.5$, the effect of the NS5--brane instanton is negligible. For $g_s = 0.5$ one sees from the table that a volume of order $\mathcal V \simeq 15$ yields corrections to the potential that cannot be ignored in a KKLT scenario. For higher values of the string coupling constant, $g_s = 1$, the corrections to the potential at $\mathcal V \simeq 60$, are of the same order as the value of the KKLT--potential at its minimum. In that case, NS5--brane instantons cannot be ignored and can change the KKLT AdS vacuum to become dS, although fine tuning is required.

Quantum corrections (both in $\alpha'$ and $g_s$) become very important at those scales. The exact form of those corrections is not known, but we can make some rough estimates. The first correction in $\alpha'$ scales as $\hat \xi^3 / \mathcal V^4$, where $\hat \xi = \xi g_s^{-3/2}$, and higher corrections are expected to be further suppressed by factors $\hat \xi/\mathcal V$. For the values $g_s=1, \mathcal V=50, \chi \sim -300$ we find $\hat \xi \sim 0.7$ and
\begin{align*}
V_{\text{ NS5}} \sim \frac{\xi^5}{ \mathcal V^6}.
\end{align*}
This can be of the same order as next--to--subleading corrections in $\alpha'$.

\subsection{\protect{Fivebranes and orientifold projections}}

We have shown the influence of a NS5--brane instanton on the scalar potential. There is, however,
a subtlety in the microscopic string theory that needs to be addressed. The NS5--brane instanton
arises from the wrapping of the 10--dimensional NS5--brane soliton solution. In 10 dimensions, the
NS5--brane is the magnetic source of the NS--NS $B_2$ field. Such a wrapping naturally arises when
we compactify six internal dimensions. In such a
compactification, the 4--dimensional part of the $B_2$ field is dualized to an axion $\sigma$, and
the NS5--brane instanton yields exponential corrections of the form
\begin{align*}
  {\rm e}^{-\frac {\mathcal V} {g_s^2} |Q| + i Q \sigma},
\end{align*}
where $Q$ is the instanton charge. As usual, the instanton action contains an imaginary part, that distinguishes between instantons ($Q>0$) and anti--instantons ($Q<0$). Microscopically, this distinction arises when one has to specify the orientation of the wrapping relative to the orientation of the CY. However, the field $\sigma$ gets projected out in an orientifold; see e.g.~\cite{Grimm:2004ua}. In the ten--dimensional picture, this corresponds to saying that the space--time part of $B_2$ gets projected out, but the NS5--brane couples magnetically to this field. These considerations seem to lead, on the one hand, to the conclusion that an NS5--brane instanton cannot exist, at least not in the traditional sense.

On the other hand, one could argue that at the level of the effective action, all even combinations in
$\sigma$ survive the projection. Examples are $\exp(-\mathcal V/g_s^2)$ and $\exp(-\mathcal
V/g_s^2) \cos \sigma$. The last term can be interpreted as an instanton ($Q=1$) -- anti--instanton
pair ($Q=-1$). Such a pair would annihilate, unless some other mechanism stabilizes the pair. They
cannot be separated in the internal manifold as e.g. for D3--branes, because they wrap the entire CY. This suggest that they cannot preserve $N=1$ SUSY after the orientifold projection, but in the next section, we show that they still can be written in a $N=1$ supergravity action. So we are led to the conlusion that
instantons do remain present after taking the orientifold projection.

The situation can be described with the following diagram:
\begin{align}\label{diagramorientifold}
    \begin{array}{rlll}
      &\text{ II String theory}/CY & \longrightarrow  & N=2, \text{effective supergravity description} \\
      & \ &\\
      &\Big\downarrow \quad \text{ orientifold} &   & \Big\downarrow \quad \text{orientifold} \\
      & \ &\\
       &\text{ II String theory}/CY _\text {or} & \longrightarrow  &  N=1, \text{effective supergravity description} \\
    \end{array}
\end{align}

Starting from the full--fledged string theory in the top left, on can obtain an effective $N=2$ supergravity description, containing effects from NS5--brane instantons. We then orientifold by simply putting $\sigma=0$ (and other fields)
in this action and obtain an $N=1$ theory.
This procedure shows that there is a contribution from NS5--branes consistent with $N=1$ SUSY. However, one could argue that the correct way to proceed is to incorporate the orientifold projection in string theory, and then calculate the low--energy effects of a NS5--brane. It would be interesting to compare these two approaches; we leave this question open for further investigation.

The situation is better understood before the orientifold projection, when we still have $N=2$. We will now turn to this setting.

\section{\protect{The N=2 scenario}}
In this section we will describe our results in the more stringent language of $N=2$ supergravity.
In this setting we have good control over the possible quantum corrections. Furthermore, there are no
subtleties with the orientifold projection of the NS5--brane, as discussed at the end of the previous section.

Although the previous section dealt with IIB string compactifications, we will change in this section to type IIA models. The reason is of technical origin, as the dimension of the hypermultiplet moduli space in IIA is given by $4(h_{1,2} +1)$ (as opposed to $4(h_{1,1} + 1)$ for IIB), and $h_{1,2}$ can be set to zero for rigid CY's. This yields a four--dimensional moduli--space, which simplifies the analysis. Moreover, NS5--brane instantons were analyzed in these models in~\cite{Alexandrov:2006hx,Davidse:2004gg}, and we will make use of these results. We expect that the results for IIA carry over to IIB.

\subsection{Gauged and ungauged N=2 supergravity}
In ungauged $N=2$ supergravity, the moduli space has the local product structure
\begin{align}\label{product}
  \mathcal M^K \times \mathcal M^Q\,.
\end{align}
For type IIA strings, the special K\"ahler manifold $\mathcal M^K$ has dimension $2h^{1,1}$ and is spanned by the scalars
in the vector multiplets, corresponding to the deformation of the K\"ahler form. The
quaternionic-K\"ahler space $\mathcal M^Q$ is spanned by the scalars in the hypermultiplets and is
$4(h^{1,2}+1)$ dimensional.
The manifold $\mathcal M^K$ is described in terms of a prepotential $F(X^I)$, where
$I=0,\ldots,h^{1,1}$. In supergravity, this can be any holomorphic function of the $X^I$ variables
of degree two. The prepotentials obtained from IIA string theory have the specific form
\begin{align}\label{prepot}
  F(X) = \frac 1 {3!}  \frac{d_{ijk} X^iX^jX^k}{X^0} + \frac {i}{2} \zeta(3)\chi(CY) X^0X^0 -
i \sum_{k_a} n_{k_a} \Li_3({\rm e}^{2\pi i k_a X^a/X^0})\,,
\end{align}
where the first term is a tree--level contribution, the second is a perturbative one-loop correction
and the last terms are the nonperturbative worldsheet instanton contributions. Note that there is
only one perturbative correction, so the perturbative regime is under complete control. The
geometry of $\mathcal M^Q$ is known at tree--level and at one--loop. It is argued in~\cite{RoblesLlana:2006ez}
that higher loop corrections can be absorbed into field redefinitions, and if so, the entire
perturbative corrected geometry is known~\cite{RoblesLlana:2006ez, Antoniadis:2003sw, Anguelova:2004sj}. If we
restrict ourselves to a rigid CY manifold, which has $h^{2,1}=0$ by definition, there is only one
hypermultiplet, which is called the universal hypermultiplet (UHM).

Gauged supergravities arise when isometries on the moduli space are gauged. They give rise to
scalar potentials that are consistent with $N=2$ supersymmetry. Microscopically, gauged
supergravities arise when fluxes (in the RR and NS--NS sector) are turned on.
For the purpose of
our paper, it suffices to look at abelian isometry groups.

The scalar potential is determined by the geometrical data of the moduli space, such as the choice
of killing vectors $k_I$ and their corresponding moment maps $\vec \mu_I$. For further details on
the gauging, we refer to appendix~\ref{ap:n2} and references therein. The result for the scalar
potential is
\begin{align}\label{generalpotential}
V =&-4  \left[2 G_{\alpha \bar \beta} k^{\alpha}_I k^{\beta}_J + 3 \vec \mu_I \cdot \vec
\mu_J\right]\!\! \frac {X^I \bar X^J}{N_{MN} X^M \bar X^N}-4 N_{MN} X^M \bar X^N \mathcal
M_{IJ}  N^{IK}N^{JL} \vec \mu_K \cdot \vec \mu_L\,.
\end{align}
In this formula, $G_{\alpha \beta}$ is the metric on the hypermultiplet space. The gauged
isometries are represented by the Killing vectors $k^\alpha_I$ and their moment maps $\vec \mu_I$.
The matrices $N_{MN}$ and $\mathcal M_{IJ}$ are
defined by
\begin{align*}
  \quad N_{IJ} &= -i F_{IJ} + i \bar F_{IJ},\\
  \mathcal M_{IJ} &= \frac 1 {[N_{MN}X^M\bar X^N]^2} \left[ N_{IJ} N_{KL} - N_{IK} N_{JL} \right]
\bar X^K X^L\,,
\end{align*}
where $F_I = \partial_I F$ etc. In our conventions, both $G_{\alpha \bar \beta}$ and $\mathcal M_{IJ}$ are negative definite, so
the first term is a positive contribution. The second is negative, whereas the last one is
positive. There is an additional term in non-abelian gaugings which can be omitted for our analysis.

\subsection{Including NS5--brane corrections}
We will now make those expressions explicit for the UHM. The perturbatively corrected metric $G_{\alpha \beta}$ on the
UHM space is given by~\cite{Antoniadis:2003sw}
\begin{align}
  {\rm ds}_\text {UHM}^2 = \frac {r+2c} {r^2(r+c)} {\rm d}r^2+  \frac {r+2c}{r^2} ({\rm d}\chi^2+{\rm d}\varphi^2)+\frac {r+c}{r^2(r+2c)}({\rm d}\sigma + \chi {\rm d} \varphi)^2.
\end{align}
The four bosonic fields are an axion $\sigma$ from the dualization of the NSNS twoform $B_2$, two
RR scalars $\chi,\varphi$ and the four--dimensional dilaton $g_4$
\begin{align}\label{dilatons}
  r \equiv {\rm e}^{\phi_4} = \frac 1 {g_4^2} = \frac {\mathcal {V}}{g_s^2}\,.
\end{align}
The relation between the four--dimensional string coupling constant $g_4$ and the
ten--di\-men\-sio\-nal string coupling constant $g_s$ is important, as it will introduce factors
of the volume into our future expressions.

The constant $c$ encodes the one--loop correction and is proportional to the Euler number
\begin{align*}
  c = -\frac{\chi(CY)}{12 \pi} = - \frac{h^{1,1}}{6\pi}\,,
\end{align*}
where the second equality holds on a rigid CY. The contributions from a single NS5--brane instanton
to the UHM metric have been derived in~\cite{Davidse:2004gg, Alexandrov:2006hx}. To leading order in the
semiclassical approximation, the metric reads
\begin{multline}\label{metricUHM}
  {\rm ds}_\text {UHM}^2 = \frac {r+2c} {r^2(r+c)} {\rm d}r^2+  \frac {r+2c}{r^2} (1-Y){\rm d}\chi^2+\frac {r+2c}{r^2} (1+Y){\rm d}\varphi^2\\+\frac 2 r \widetilde Y {\rm d}\chi {\rm d}\varphi +
  \frac {r+c}{r^2(r+2c)}({\rm d}\sigma + \chi {\rm d} \varphi)^2.
\end{multline}
The quantities $Y$ and $\widetilde Y$ are defined as\footnote{Compared to~\cite{Davidse:2004gg} we
have taken $\chi_0 =0$. Its dependence can easily be restored.}
\begin{align*}
  Y = 4 C (2 \chi^2-1) r^{-1-c} \cos (\sigma) {\rm e}^{-r-\frac 12 \chi^2-c},\quad
  \widetilde Y = 4 C (2 \chi^2-1) r^{-1-c} \sin (\sigma) {\rm e}^{-r-\frac 12 \chi^2-c}.
\end{align*}
The factor $C$ is a numerical constant which could not be determined. This solution has a shift symmetry associated with $\varphi$ which we can gauge, using the
graviphoton as gauge field. The field $\varphi$ is obtained by expanding the 10--dimensional RR field $\widehat C_3$ over one of the $2 (h^{2,1}+1)=2$ cycles in $H^3$; gauging the isometry associated with $\varphi$ has a microscopic interpretation of adding NS flux over this cycle~\cite{Louis:2002ny}. Moreover, the shift symmetry is not broken by NS5-brane instantons, so there is no obstruction in
gauging this isometry by fluxes in the presence of instantons
\cite{Kashani-Poor:2005si,Anguelova:2006sj}.

The gauging of this isometry leads to a scalar potential of the type given in~\eqref{generalpotential}, with a Killing vector $k = \partial_\varphi$.
Upon inserting the prepotential~\eqref{prepot} without the worldsheet instantons, one finds that the moment maps drop out of the equation for the
potential~\eqref{generalpotential}, and only the norm of the Killing vector remains. The only dependence on the vector multiplet moduli comes from the factor $(N_{MN}X^MX^{\bar
N})^{-1}$. The details of this calculation can be found in appendix~\ref{ap:n2}.

Without the worldsheet instanton corrections, we then find the scalar potential
\begin{align}\label{VUHM}
V &= \frac{2}{4\mathcal V + e} \left[ -2 G_{\alpha \bar \beta} k^\alpha k^\beta \right]\nonumber\\
&= \frac{4}{4\mathcal V + e} \left( \frac{ 4 ( r+2c)^2+4 (r+c)\chi^2}{r^2 (r+2c)}+ 16 \, C {\rm e}^{-c-r-\chi^2/2}
r^{-2-c} (2\chi^2-1) \cos (\sigma) \right),
\end{align}
where $e = \frac 12 \zeta(3)\chi(CY)$ and $C$ is the undetermined overall constant. Reinstating
all volume factor dependencies using~\eqref{dilatons}, this has the schematic form
\begin{align*}
  V &= V_0 + \widetilde C \mathcal V^{-3-c} {\rm e}^{-\mathcal V/g_s^2},
\end{align*}
where $\widetilde C = 16 C g_s^{4+2c} {\rm e}^{-c-\chi^2/2} (2\chi^2-1) \cos (\sigma) $. We also neglect the correction due to $e$ in the 2nd term, because it is subleading.

We see how a NS5--brane contribution can be included into $N=2$ type IIA supergravity. It would be interesting to repeat this exercise including the worldsheet instantons.

We now truncate this theory to obtain an $N=1$ description.

\subsection{Truncation to $N=1$}
To clarify the relation to the previous section, we will now perform a truncation of this theory. We make an orientifold inspired truncation to $N=1$ at the level of the effective action (see figure~\eqref{diagramorientifold}). A similar truncation has been done in~\cite{Balasubramanian:2004uy}. We follow the orientifold rules from~\cite{Grimm:2004ua}. Because we merely truncate the theory, there should still be a local product structure as in ~\eqref{product}, but now the product is between two K\"ahler manifolds. Furthermore, for simplicity we restrict ourself to the cubic prepotential and therefore put $e=0$.

The universal hypermultiplet loses half of its fields under truncation to become a chiral $N=1$ multiplet. We keep the four--dimensional dilaton $r$ and project out the axion $\sigma$. From the RR scalars $\chi,\varphi$ we can choose which we keep. We gauged the isometry on $\varphi$, which corresponds to a NS--flux on the cycle of $\varphi$. The relevant part of the expansion of $\widehat H_3$ and $\widehat C_3$ is given by
\begin{align}
\widehat C_3 &= \chi \alpha + \varphi \beta\\
\widehat H_3 &= p \, \alpha + q \, \beta\,.
\end{align}
The field $\widehat C_3$ is expanded over a basis of the third cohomology group $H^3$, given by three--forms $\alpha,\beta$, which give the four--dimensional fields $\chi,\varphi$. The flux of $\widehat B_2$ is likewise expanded, with flux parameters $p,q$.  Under an orientifold, the RR form $\widehat C_3$ and the NS-NS flux $\widehat H_3 = {\rm d}\widehat B_2$ are even and odd respectively. We gauge the isometry associated with $\varphi$, so we want to keep the flux parameter $q$. This implies that $\beta$ should be an odd cycle. In the expansion of the even form $\widehat C_3$ we only keep even forms, and hence $\varphi$ gets projected out.

The metric then truncates to
\begin{align}\label{metricUHMtruncated}
  {\rm ds}_\text {UHM}^2 = \frac {r+2c} {r^2(r+c)} {\rm d}r^2+  \frac {r+2c}{r^2} (1-Y){\rm d}\chi^2,
\end{align}
where we have put $\sigma =0 $ in $Y$. The perturbatively corrected scalar potential is, with $e=0$
\begin{align}\label{eq:potpertn=1}
V_0 = \frac 4 {\mathcal V} \frac{(r+2c)^2 + (r+c)\chi^2}{r^2(r+2c)}\,,
\end{align}
and the NS5--brane instantons yields equation~\eqref{VUHM} with $\sigma=0$
\begin{align}
V_{\rm NS5} = \frac{16}{\mathcal V} C {\rm e}^{-c-r-\chi^2/2} r^{-2-c} (2\chi^2-1)\,,
\end{align}
where $C$ is independent of vector multiplet scalars.

We want to express these quantities in terms of a K\"ahler and superpotential. The K\"ahler potential $K$ is a sum of the K\"ahler potential $K_{\rm k}$ for the truncated K\"ahler moduli and a potential $K_{\rm Q}$ for the truncated universal hypermultiplet. In the K\"ahler sector, we have the K\"ahler potential~\cite{Grimm:2004ua}
\begin{align*}
K_{\rm k} = - \ln (\mathcal V)\ ,
\end{align*}
which follows from the choice of the cubic prepotential we made in the $N=2$ calculation, earlier in this section. The important property of this K\"ahler potential is its no--scale structure $K_{\rm k}^{i \bar \jmath} K^{\rm k}_i K^{\rm k}_{\bar \jmath} = 3$.

For the scalar potential we use expression~\eqref{scalarpotential}
\begin{align*}
V={\rm e}^K\Big(K^{\alpha\bar \beta}D_\alpha WD_{\bar \beta}{\overline W}-3|W|^2 \Big) = \frac 1 {\mathcal V} \,{\rm e}^{K_{\rm Q}} \big(K_{\rm Q}^{z \bar z} D_z W \overline {D_z W}\big)\,,
\end{align*}
where the no--scale structure in directions orthogonal to the truncated universal hypermultiplet has been used.

The perturbative part of the metric~\eqref{metricUHMtruncated} and the potential~\eqref{eq:potpertn=1} are now exactly reproduced by the K\"ahler potential and superpotential~\cite{Saueressig:2005es}
\begin{align*}
  K_{\rm Q} &= -2 \ln \left[ (z+\bar z)^2 - 16 c \right]\,\!,\\
  W &= 16 z\,.
\end{align*}
These are formulated in terms of the chiral field $z$ defined by
\begin{align*}
  z = 2 \sqrt{r+c} + i \chi\,.
\end{align*}
We use conventions for which ${\rm ds}^2 = 4 K_{z \bar z} {\rm d}z {\rm d}\bar z$, as in~\cite{Saueressig:2005es}.

We now also want to describe the NS5--brane instanton contribution. This scales as
\begin{align*}
  \exp\left(-r-\frac {\chi^2} 2-c\right) = \exp\left(\frac 1 {16}(z^2-6 z \bar z + \bar z^2)\right),
\end{align*}
which is not holomorphic in $z$. Therefore, we cannot correct the superpotential with such a term. The
correction will take place in the K\"ahler potential and in the definition of the $N=1$ chiral
field. Both the metric~\eqref{metricUHMtruncated} and the potential~\eqref{VUHM} are reproduced up
to leading order by the chiral field and the K\"ahler potential
\begin{align}\label{eq:KenZNS5}
z &= 2 \sqrt{r+c} + i \chi + 2 C {\rm e}^{-r-\frac 12 \chi^2-c} r^{-2-c} \left( \sqrt r (1-2\chi^2)   + i \chi (2\chi^2-5) \right),\\
K_{\rm Q} &= -2 \ln \left[(z+\bar z)^2 - 16c \right] + C \exp
\left[ \frac 1 {16}(z^2-6 z \bar z + \bar z^2) \right] 4^{9+2c}  \frac{ (z+\bar z)^{-4-2c} (1+\frac 12
(z-\bar z)^2)}{( z - 3 \bar z)(  \bar z-3z)} \nonumber \,,\\
W &= 16z \nonumber\,.
\end{align}
Because our four--dimensional dilaton $r$ contains a factor of the volume $\mathcal V$, the leading term in the K\"ahler potential is equal to
\begin{align}\label{eq:KNS5leading}
  K_{\rm Q} = K_0 + B \mathcal V^n {\rm e}^{-\mathcal V/g_s^2}\,,
\end{align}
where we have defined
\begin{align}\label{B-Ctilde}
B = 64 C {\rm e}^{-c-\chi^2/2} (1-2\chi^2) g_s^{-2n},\quad n=-3-c.
\end{align}

The overall factor $B$ can depend on other moduli. In this setting, the leading dependence on $\chi$ is explicit in~\eqref{B-Ctilde}. We expect~\eqref{eq:KNS5leading} to hold also for non--rigid CY with $h_{1,2} \neq 0$. In that case the factor $B$ presumably depends on the other hypermultiplet scalars.

This confirms our proposal of~\eqref{KIIBNS5} and~\eqref{eq:valueofn}, where we apply it to type IIB string theory.
\newpage
\subsection*{Acknowledgements}
We would like to thank Lilia Anguelova,  Tim Baarslag, Ralph Blumenhagen, Jan de Boer and Frank Saueressig for discussions and correspondence. This work is partially supported by the European Union RTN network MRTN-CT-2004-005104 and INTAS contract 03-51-6346.

\appendix
\section{Calculations in $N=1$}\label{ap:ns5calc}
In this appendix we give some details of the calculations which have been used in section 2.

We want to calculate the derivatives and inverses for the K\"ahler potentials~\eqref{KIIBLVS}
\begin{align}\label{ap:kahlerpot}
  K_0 &= - \ln (-i (\tau-\bar \tau)) - 2 \ln \left(\mathcal V + \frac {\hat \xi}2\right),\\
  K_{\rm NS5} &= B \mathcal V^n \exp\big(- \mathcal V \tau_2^2)\,,
\end{align}
where we use
\begin{align*}
  \tau &= l + i {\rm e}^{\phi},\quad \tau_2 = \text {Im} \tau\,,\\
 6 \mathcal V &= d_{ijk} t^it^jt^k = d_{ij}t^it^j = d_i t^i\,,\\
 \hat \xi &= \xi (\text {Im} \, \tau)^{3/2}.
\end{align*}
We introduce $d^{ij}$ as the inverse of $d_{ij}$, and denote $\mathcal A := \mathcal V + \hat \xi
/2$. We first consider the case where $G^a = 0$. From $T_i = \tau_i + ib_i$ (equation~\eqref{chiral-fields1} for $G^a=0$) we find
\begin{align*}
  \frac {\partial t^j}{\partial T_i} = \frac 14 d^{ij},\quad \frac {\partial \mathcal V}{\partial T_i} = \frac 18 t^i.
\end{align*}
We can then calculate
\begin{align*}
K^0_{T_i} &= - \frac 1 4 \frac {t^i}{\mathcal A},\quad K^0_\tau = i \tau_2^{-1} \left(\frac {3 \hat \xi}{4\mathcal A} + \frac 12 \right)\,,\\
K^0_{T_i\bar T_j} &= \frac {G^{ij}}{\mathcal A^2},\quad
G^{ij} = - \frac 1 {16} \mathcal A d^{ij} + \frac 1 {32} t^it^j\,,\\
K^0_{i \bar \tau} &= \frac {3i}{32\mathcal A^2} \tau_2^{-1} \hat \xi t^i\,,\\
K^0_{\tau \bar \tau} &= \frac 1 {16} \tau_2^{-2} \mathcal A^{-2} (4 \mathcal V^2 + \mathcal V \hat
\xi + 4 \hat \xi^2)\,.
\end{align*}
This can be inverted to give
\begin{align*}
K_0^{\tau \bar \tau} &= \frac{4\mathcal V - \hat \xi}{\mathcal V - \hat \xi} \tau_2^2\,,\\
K_0^{i\bar \tau} &= -\frac {3i \hat \xi}{\mathcal V - \hat \xi} \tau_2 d_i\,,\\
K_0^{i\bar \jmath} &= -8 (2\mathcal V + \hat \xi) d_{ij} + \frac {4\mathcal V-\hat \xi}{\mathcal
V-\hat \xi} d_id_j\,.
\end{align*}
We then find a familiar result, which leads directly to~\eqref{VLVS}:
\begin{align}\label{pot:bbhl}
  K_0^{\alpha \bar \beta} K^0_{\alpha} K^0_{\bar \beta} = 3 + \frac { 3\hat \xi (\mathcal V^2 + 7
  \mathcal V \hat \xi + \hat \xi^2)}{(\mathcal V - \hat \xi)(2 \mathcal V + \hat\xi)^2}\,.
\end{align}
For the NS5--brane contribution we obtain (we write $K_5 = K^5 = K_{\rm NS5}$)
\begin{align*}
K^5_{\tau \bar \tau} &= -\frac 14 K_5 \mathcal V ((\tau-\bar \tau)^2 \mathcal V + 2)\,,\\
K^5_{i \bar \tau} &= \frac i 8 B\exp(-\mathcal V\tau_2^2) \mathcal V^n (-\mathcal V \tau_2 + n + 1) \tau_2 t^i\,,\\
K^5_{i \bar \jmath} &= \frac 1 {64} B t^it^j \mathcal V^{n-2} \left(\mathcal V^2 \tau_2^4
-\mathcal V 2 n \tau_2^2 + (n-1)n\right)\exp(-\mathcal V \tau_2^2)\,.
\end{align*}
We are interested in the leading term in the potential, so we want to investigate the powers of the
volume. If we denote volume powers with $[\cdot]$, then
\begin{align*}
  [\mathcal V] = 1, \quad [d_{ijk}]=0, \quad [t^i]=\frac 13, \quad [d_{ij}]=\frac 13\,.
\end{align*}
All the one--instanton terms are multiplied by $\exp (-\mathcal V \tau_2^2)$. To determine the leading term, we have to find the highest power of the volume $\mathcal V$ in the polynomial which appears in front of this exponent. Therefore, we do not include the factor $\exp(-\mathcal V \tau_2^2)$ in the counting, or equivalently we put $[\exp(-\mathcal V \tau_2^2)]=0$.

The various terms have the following leading volume dependencies:
\begin{align*}
  [K_0^{i\bar \jmath}] &= 4/3 &&& [K^5_{i    \bar \jmath}] &= n+2/3  &&& [K_0^{i \bar \jmath} K^5_{\bar \jmath k   } K_0^{\bar k    l}] &= n+10/3\\
  [K_0^{i\bar \tau}] &= -1/3  &&& [K^5_{i    \bar \tau  }] &= n+4/3  &&& [K_0^{i \bar \tau}   K^5_{\bar \tau   k   } K_0^{\bar k    l}] &= n+7/3\\
  [K_0^{\tau\bar \tau}] &= 0  &&& [K^5_{\tau \bar \tau  }] &= n+2    &&& [K_0^{i \bar \tau} K^5_{\bar \tau   \tau} K_0^{\bar \tau l}] &= n+4/3\,.
\end{align*}
The leading contribution is given by (we use $\simeq$ here to denote equality up to subleading terms)
\begin{align*}
K_5^{i\bar \jmath} = K_0^{i\bar \alpha}K^5_{\bar \alpha \beta}K_0^{\beta \bar \jmath} \simeq K_0^{i\bar l}K^5_{\bar l k}K_0^{k \bar \jmath} \simeq B  \mathcal V^{n+2} g_s^{-4}
\exp(-\mathcal V/g_s^2)d_id_j\,.
\end{align*}
In the potential we find then $[K_5^{i\bar \jmath} |\partial_i K_0|^2] = (n+10/3) - 2/3 - 2/3 =
n+2$. The other term is $[K_0^{ij} \partial_i K^5 \partial_{\bar \jmath} K_0] = 4/3 + (n+1/3) - 2/3
= n+1$, which is subleading with respect to the terms above. Then the leading contribution to the scalar potential is given by
\begin{align}\label{ap:potfinal}
  V &\simeq -{\rm e}^{K_0}K_{{\rm NS5}}^{i\bar \jmath}\,|\partial_i K_0|^2 |W_0|^2\\\nonumber
  &= -\frac 9 8 B \mathcal V^n g_s^{-3} |W_0|^2 \exp (-\mathcal V/ g_s^2).
\end{align}

Let us now consider the effects of non-zero $G^a$, to clarify the statements made after equation~\eqref{ns5inverse}. From the definition~\eqref{chiral-fields1}
\begin{align*}
  T_i =  \tau_i + ib_i + \frac{i}{\tau-\bar\tau} \, d_{iab}G^a (G-\bar G)^c\,,
\end{align*}
we find that
\begin{align*}
  \frac {\partial t^j}{\partial T_i} = \frac 14 d^{ij},\quad
  \frac {\partial{t^i}}{\partial{G^a}} = \frac 1 4 d^{i j} d_{jab}b^b,\quad
  \frac {\partial{t^i}}{\partial \tau} = \frac i 2 d_{iab}b^ab^b,
\end{align*}
and hence
\begin{align*}
  \frac {\partial \mathcal V}{\partial T_i} = \frac 18 t^i,\quad
  \frac {\partial \mathcal V}{\partial{G^a}} = \frac 12 t^j d_{jac}b^c = \frac 12 d_{ac}b^c,\quad
   \frac {\partial \mathcal V}{\partial \tau} = \frac i 4 t^j d_{jab}b^ab^b = \frac i 4 d_{ab}b^ab^b.
\end{align*}
In the last two expressions the factor $t^j$ is bound with the factor $d_{jac}$ and cannot combine with a $d_j$ to
form a power of the volume. 

The expression for~\eqref{ns5inverse} also contains inverse metrics. If we use the expressions
for the tree--level K\"ahler metric in~\cite{Grimm:2004uq}, we can explicitly determine the volume
dependence, and we find
\begin{align*}
  K_0^{ia}K^5_{ab}K_0^{bj} t^it^j &\sim \mathcal V^{n+2}\\
  K_0^{ik}K^5_{kl}K_0^{lj} t^it^j &\sim \mathcal V^{n+4}\\
  K_0^{ia}K^5_{ak}K_0^{kj} t^it^j &\sim \mathcal V^{n+3},
\end{align*}
and the leading term does not contain the fields $G^a$. We do not know if this property holds when
we include quantum corrections to the K\"ahler potential, but as quantum corrections are expected
to be subleading in the volume, we expect this to be the case.

\section{\protect{Potentials in $N=2$}}\label{ap:n2}

In this appendix we give some more details of the calculation in the $N=2$ setting. This
appendix derives a general form of the scalar potential. The next appendix specializes this to the
UHM and the Przanowski metric.

We use the formalism from~\cite{deWit:2001bk} with the vector prepotential
\begin{align}\label{prepot-pert}
  F = \frac 1 {3!}  \frac{d_{ijk} X^iX^jX^k}{X^0} + \frac {i}{2} \zeta(3)\chi(CY)
  X^0X^0.
\end{align}

{}From the prepotential we define
\begin{align*}
N_{IJ} &= -i F_{IJ} + i\bar F_{IJ} = 2 {\rm Im} \, F_{IJ},\\
\mathcal M_{IJ} &= \frac 1 {[N_{MN}X^M\bar X^N]^2} \left[ N_{IJ} N_{KL} - N_{IK} N_{JL} \right]
\bar X^K X^L\,,
\end{align*}
and then the scalar potential is given by
\begin{align*}
V =&-4 g^2 \left[2 G_{\alpha \bar \beta} k^{\alpha}_I k^{\beta}_J + 3 \vec \mu_I \cdot \vec
\mu_J\right] \frac {X^I \bar X^J}{N_{MN} X^M \bar X^N} \\ &- g^2 N_{MN} X^M \bar X^N \mathcal
M_{IJ} \left[4 N^{IK}N^{JL} \vec \mu_L \cdot \vec \mu_L - \frac {f_{KL}^I X^K \bar X^L}{N_{PQ} X^P
\bar X^Q} \frac {f_{MN}^J X^M \bar X^N}{N_{PQ} X^P \bar X^Q}\right],
\end{align*}
where $g$ is an overall factor to make the terms which are a result of the gauging more explicit in the Lagrangian; we put $g=1$ from now on. In general, each vector field in the vector multiplets can be used to gauge one of the $h^{1,1}+1$ different killing vectors $k^I$. In our setting, there is only one isometry $k$, which we gauge by the graviphoton. The index $I$
therefore only attains the value $0$. If we use $\vec \mu_I = \delta_I^0 \vec \mu$ we obtain
\begin{align*}
V =&- \frac{4}{N_{MN} X^M \bar X^N } \Big( \left[2 G_{\alpha \bar \beta} k^{\alpha} k^{\beta} +
3 \vec \mu ^2\right] X^0 \bar X^0 + ( N_{KL} N^{00} - \delta_K^0 \delta_L^0) \bar X^K X^L  \vec
\mu^2 \Big),
\end{align*}
and the term depending on $f_{KL}^I$ is zero for abelian gaugings. We now use the prepotential
\begin{align*}
F = \frac 1 {3!} d_{ijk} \frac {X^iX^jX^k}{X^0} + \frac 12 e X^0 X^0,
\end{align*}
where $e = \frac i2 \zeta(3) \chi(CY)$ is purely imaginary. Using $X^i/X^0 =
z^i = b^i + i t^i$, we find
\begin{align*}
F_{00} &= \frac 13 d_{ijk} z^iz^jz^k + e & {\rm Im}\, F_{00} &= d_{ijk} b^ib^jt^k - \frac 13 d_{ijk}
t^it^jt^k\\
 F_{0i} &= -\frac 12 d_{ijk} z^jz^k & {\rm Im}\,  F_{0i} &= -d_{ijk} b^jt^k\\
 F_{ij} &=  d_{ijk} z^k & {\rm Im}\,  F_{ij} &=  d_{ijk} t^k\,.
\end{align*}
Using the abbreviations $d_{ij} = d_{ijk} t^k, d_i = d_{ij}t^j, 6 \mathcal V = d_{ijk}t^it^jt^k$ we
find
\begin{align}
N_{00} &= 2 d_{ij} b^ib^j -4\mathcal V' & N^{00} &= \frac {-1}  {4\mathcal V'}\nonumber
\\
N_{0i} &= -2 d_{ij} b^j & N^{0i} &= \frac {-b^i} {4\mathcal V'}\\
N_{ij} &= 2d_{ij} & N^{ij} &=  \frac{-b^ib^j}{4\mathcal V'} + \frac 12 d^{ij}\nonumber \\
{\rm e}^{-K} &\equiv N_{IJ}X^I \bar X^J = (8 \mathcal V + 2 e) X^0 \bar X^0 \,.\nonumber
\end{align}
where we have written $\mathcal V' := \mathcal V - \frac 12 e$. For the scalar potential we then finally find equation~\eqref{VUHM}
\begin{align}\label{V-cubic}
V&=- \frac{4}{8 \mathcal V + 2 e} \left( \left[2 G_{\alpha \bar \beta} k^{\alpha}
k^{\beta} + 3 \vec \mu ^2\right]  + ( 8 \mathcal V' \frac {-1} {4\mathcal V'} - 1) \vec \mu^2 \right), \nonumber\\
&= \frac{2}{4\mathcal V + e} \left[ -2 G_{\alpha \bar \beta} k^\alpha k^\beta \right].
\end{align}
The moment maps drop out and the scalar potential is positive definite.

\section{The Przanowski metric}
In this appendix we repeat some of the results of~\cite{przanowski,Alexandrov:2006hx}, which are used to determine the NS5--brane one--instanton corrected $N=2$ moduli space in section 3.2.

In \cite{przanowski}, it has been shown that a four-dimensional quaternionic-K\"ahler manifold $M$
can be described in terms of a partial differential equation for a single, real function. Locally, the metric takes the form
\begin{align}\label{eq:prmetric}
g &= g_{\alpha \bar \beta} (dz^\alpha \otimes dz^{\bar \beta} +
dz^{\bar \beta} \otimes dz^{\alpha})\\
&= g_{1\b1} dz^1 dz^\be + g_{1\b2} dz^1 dz^\bt + g_{2 \be} dz^2 dz^\be + g_{2 \bt} dz^2dz^\bt +
c.c.,\nonumber
\end{align}
where indices $\alpha,\beta,\bar \alpha,\bar \beta = 1,2$, and we have used the usual convention of
complex conjugation $z^{\bar \alpha} := \overline {z^\alpha}$. The Hermicity of this metric is
encoded in the requirement $\overline {g_{\alpha \bar \beta}} = g_{\beta \bar \alpha}$. The
elements $g_{\alpha \bar \beta}$ are now defined in terms of a real function $h =
h(z^\alpha,z^{\bar \alpha})$ via
\begin{align}\label{eq:prmetric2}
g_{\alpha \beta} = 2 \left(h_{\alpha \bar \beta} + 2 \delta_{\alpha}^2 \delta_{\bar \beta}^2 {\rm e}^h \right),
\end{align}
where the subscript $\alpha$ on $h_{\alpha}$ indicates differentiation of the function with respect
to $z^\alpha$. We have changed the sign of our defining function $h$ with respect to the
original function $u$ used by Przanowski, as it offers a slightly more convenient form to work
with.

The differential equation which determines the function $h$ is the non--linear partial differential
equation
\begin{align}\label{eq:prz}
h_{1\b1} h_{2\bt} - h_{1\bt} h_{\be2} + (2 h_{1\be} - h_1 h_{\be}) {\rm e}^h = 0\,.
\end{align}

\subsection{Solutions to the master equation} The equation~\eqref{eq:prz}
is a difficult partial differential equation. There have been various approaches in the literature
which found exact and approximate solutions to the master equation. By imposing
additional symmetries on the manifold $M$, one can simplify the master equation. Imposing one
isometry reduces this equation to the Toda equation~\cite{przanowski}. Upon imposing two commuting isometries one
obtains the Calderbank-Pedersen metrics~\cite{calderbank-2002-60}.

In~\cite{Alexandrov:2006hx}, solutions to the master equation where obtained which corresponded to
NS5--brane instantons. The relation between the complex coordinates and the real coordinates is
given by
\begin{align*}
  z^1 = \frac 12 (u+i\sigma),\quad z^2 = \frac 12 (\chi + i\varphi), \quad u \equiv r - \frac 12 \chi^2 +
  c \log (r+c)\,.
\end{align*}
The leading term of the one--instanton contribution is captured by
\begin{align}\label{equation:h}
  h &= h_0 + \Lambda,\quad h_0 = \log (r+c) - 2 \log r\\
  \Lambda &= C r^{-2-c} \cos(\sigma) \exp\left[-r + \frac 12 \chi^2\right].\nonumber
\end{align}
{}From the metric we only need the length of the Killing vector $k = \partial_\varphi$, which can be found from~\eqref{eq:prmetric},~\eqref{eq:prmetric2} and~\eqref{equation:h} and is given by
\begin{align*}
  -G_{\alpha \beta} k^\alpha k^\beta = \frac { 4((r+2c)^2 + (r+c)\chi^2)}{r^2(r+2c)} + 16 C r^{-2-c}
  (2\chi^2-1)
  \exp(-c-r-\chi^2/2)\,.
\end{align*}
Inserting this into~\eqref{V-cubic} yields~\eqref{VUHM}.

\subsection{Moment maps}
Although the moment maps are not present in the scalar potential, we include their calculation for
completeness. We follow the conventions on quaternionic-K\"ahler geometry from~\cite{Davidse:2005ef}.

We want to find vielbeins $a, b$ for the metric~\eqref{eq:prmetric} such that
\begin{align*}
a \otimes \bar a + b \otimes \bar b + c.c. = {\rm ds}^2.
\end{align*}
Using the Ansatz $a = \alpha {\rm d}z^1 + \beta {\rm d}z^2, b = \gamma {\rm d}z^1 + \delta {\rm d}z^2$ we find
\begin{align*}
a &=\sqrt{2 h_{1\b1}}\, {\rm d}z^1 + \sqrt {2} \frac{ h_{\be2}}{\sqrt{h_{1\b1}}}\,
{\rm d}z^2, \\
b &= \sqrt{2} {\rm e}^{h/2} \sqrt{\frac{h_1h_\be}{h_{1\b1}}} \, \,{\rm d}z^2.
\end{align*}
{}From those, we determine the $SU(2)$ connection one--forms
\begin{align*}
\omega^1 &= i \frac{{\rm e}^{h/2}}{\sqrt{h_1h_\be}} (h_\be {\rm d}z^2 - h_1
{\rm d}z^\bt),\\
\omega^2 &= - \frac{{\rm e}^{h/2}}{\sqrt{h_1h_\be}} (h_\be {\rm d}z^2 + h_1
{\rm d}z^\bt),\\
\omega^3 &= -\frac i 2 \left(h_1 - \frac{h_{1\be}}{h_\be} +
\frac{h_{11}}{h_1} \right) {\rm d}z^1\\
&\phantom{= }-\frac i2 \left(h_2- \frac{h_{\be2}}{h_\be} + \frac{h_{12}}{h_1} \right) {\rm d}z^2 + c.c.
\end{align*}
As a non-trivial check, we can use the tree-level UHM metric, and these one-forms agree with the
those obtained in~\cite{Davidse:2005ef}. Notice that the situation drastically simplifies when
there is an additional killing vector in the direction $i(\partial_1 -
\partial_{\bar 1})$, because then $h_1 = h_{\bar 1}$.

We now gauge the isometry associated with $\varphi$. In the complex coordinates, this is the vector
\begin{align*}
k = \frac 12 i(\partial_2 - \partial_\bt),
\end{align*}
where the normalization is such that $k = \partial_\varphi$. Calculations of the moment maps is now
straight-forward and after some algebra we find
\begin{align*}
\vec \mu =
\left(
  \begin{array}{l}
  \frac{{\rm e}^{h/2}} { \sqrt  {h_1 h_\be}} (h_1 + h_\be)\\
 -i \frac{{\rm e}^{h/2}} { \sqrt  {h_1 h_\be}} (h_1 - h_\be)\\
  - h_2
  \end{array}
\right),
\end{align*}
which are real ($h_2 = h_\bt$).

The square of the moment maps therefore reads
\begin{align*}
\vec \mu^2&= (4{\rm e}^h + h_2^2) = 4 {\rm e}^h + (\partial_\chi h)^2,
\end{align*}
where we have used $\partial_\varphi h =0$. This last expression is valid in the coordinates
($u,\sigma,\chi,\varphi$). Changing to the coordinates ($r,\sigma,\chi,\varphi$) amounts to
changing the derivatives according to
\begin{align*}
\partial_\chi \rightarrow \partial_\chi + \chi \frac{r+c}{r+2c}
\partial_r.
\end{align*}

\end{document}